\documentclass[a4paper,11pt]{article}
\usepackage{amsfonts}
\usepackage{amsmath,amssymb,amscd}
\usepackage{appendix,marginnote,tikz,pgf,mathtools}
\usepackage{comment}
\baselineskip=\normalbaselineskip

\def\ll{\left\langle}
\def\rr{\right\rangle}
\def\be{\begin{equation}}
\def\ee{\end{equation}}
\def\bea#1\een{\begin{align}#1\end{align}}
\def\p{\partial}
\def\matt[#1,#2,#3,#4]{
	\left(\begin{array}{cc} #1 & #2 \\ #3 & #4 \end{array}\right)}
\def\mat3[#1,#2,#3,#4,#5,#6,#7,#8,#9]{
	\left(\begin{array}{ccc}
	#1 & #2 & #3 \\ #4 & #5 & #6 \\ #7 & #8 & #9
	\end{array}\right)}
\begin{document}
\setcounter{footnote}{0}
\setcounter{tocdepth}{3}
\bigskip
\def\thefootnote{\arabic{footnote}}
\begin{titlepage}
\renewcommand{\thefootnote}{\fnsymbol{footnote}}
\begin{normalsize}
\begin{flushright}
\begin{tabular}{l}
UTHEP-737
\end{tabular}
\end{flushright}
  \end{normalsize}

~~\\

\vspace*{0cm}
    \begin{Large}
       \begin{center}
         {The null identities for boundary operators in the $(2,2p+1)$ minimal gravity}
       \end{center}
    \end{Large}
\vspace{0.7cm}

\begin{center}
Goro I\textsc{shiki}$^{1),2)}$\footnote[0]
            {
e-mail address : 
ishiki@het.ph.tsukuba.ac.jp},
Hisayoshi M\textsc{uraki}$^{3)}$\footnote[0]
            {
e-mail address : 
hmuraki@sogang.ac.kr}
 and
Chaiho R\textsc{im}$^{4)}$\footnote[0]
            {
e-mail address : 
rimpine@sogang.ac.kr}

\vspace{0.7cm}

     $^{ 1)}$ {\it Tomonaga Center for the History of the Universe, University of Tsukuba, }\\
               {\it Tsukuba, Ibaraki 305-8571, Japan}\\
                   
     $^{ 2)}$ {\it Graduate School of Pure and Applied Sciences, University of Tsukuba, }\\
               {\it Tsukuba, Ibaraki 305-8571, Japan}\\
               
     $^{ 3)}$ {\it Asia Pacific Center for Theoretical Physics, }\\
               {\it Pohang, Gyeongbuk 37673, Korea}\\
               
     $^{ 4)}$ {\it Department of Physics, Sogang University, }\\
               {\it Seoul 04107, Korea }\\
               \end{center}

\vspace{0.5cm}

\begin{abstract}
By using the matrix-model representation, we show that correlation numbers of
boundary changing operators (BCO) in $(2,2p+1)$ minimal Liouville gravity 
satisfy some identities, which we call the null identities. 
These identities enable us to express the correlation numbers 
of BCO in terms of those of boundary preserving operators. 
We also discuss a physical implication of the null identities as the manifestation of 
the boundary interaction.
\end{abstract}

\tableofcontents

\end{titlepage}

\section{Introduction}

The 2-dimensional gravity coupled with a minimal model of CFT has been studied 
as a good example of well-defined quantum gravitational theories 
\cite{Knizhnik1988},
which also allows a non-perturbative discrete formulation given by matrix models
\cite{BK1990, DS1990,  GM1990,Ginsparg:1991bi}.

In this paper, we  follow the one-matrix model description \cite{MSS1991, BZ2009}  
of the $(2, 2p+1)$ minimal gravity on Riemann surfaces
but focus on the description in the presence of boundaries \cite{BR2010, IR2010}. 
The boundary conditions of the minimal gravity, also referred to as 
FZZT branes \cite{FZZ2000}, 
are specified by the value of the boundary cosmological constant 
$\mu_B$ and the Kac label $(1,m)$ of the matter Cardy state.
In \cite{IR2010},  it was shown that such boundary conditions 
are realized in the matrix model by introducing a generalization of 
the resolvent operators. The disk partition function 
for the $(1,m)$ Cardy state is given by 
\begin{align}
F_m = -\langle {\rm tr} \log f_m(M) \rangle,
\label{disk with a single boundary}
\end{align}
where $f_m(M)$ is a monic polynomial of the Hermitian matrix $M$
with degree $m$ and $\langle \cdots \rangle $ stands for the expectation value 
of the one-matrix model. 
After some renormalizations, the coefficients of $f_m(M)$ are related to 
the sources of boundary operators, which preserves the $(1,m)$ boundary condition. 

One can introduce some impurities on the boundary, which interpolate 
two different boundary conditions. These are called the boundary changing operators (BCO). 
Between two boundary segments of $(1,m_1)$ and $(1,m_2)$ with different boundary 
cosmological constants, one can put a $(1,k)$ primary operator dressed by the Liouville factor $e^{\beta_k \phi}$,
where $k=|m_1-m_2|+1, |m_1-m_2|+3, \cdots, m_1+m_2-1$, 
$\beta_k =\frac{(k+1)b}{2}$ and $b^2=2/(2p+1)$.
It was shown in \cite{IR2010} that these operators are described in 
the one-matrix model as follows. 
One extends the disk partition functions to the $2\times2$ block matrix of the form
\begin{align}
F_{m_1m_2} = -\ll {\rm tr} \log \left( 
\begin{array}{cc}
f_{m_1} (M)  & g_{m_1m_2}(M) \\
g_{m_1m_2}^{\dagger}(M)  & f_{m_2}(M) \\
\end{array}
\right)  \rr.
\label{perturbed partition function}
\end{align}
Here, $g_{m_1m_2}(M)$ is a polynomial of $M$ with degree less than ${\rm min}(m_1,m_2)$.
The coefficients of $g_{m_1m_2}(M)$ provides sources of BCOs between the 
$(1,m_1)$ and $(1,m_2)$ boundaries.  Correlation numbers with more different 
boundary conditions can also be treated in the similar way by introducing 
more block structures. It was shown that this formulation 
correctly reproduces the correlation numbers of BCOs, computed in the Liouville theory approach.

In this paper, we demonstrate that the correlation numbers of BCOs satisfy 
some nontrivial identities, which we call null identities. 
The idea for deriving the null identities is the following. 
The perturbed partition function 
(\ref{perturbed partition function}) can be diagonalized to the form,
\begin{align}
F_{m_1m_2} =-\ll {\rm tr} \log \left( 
\begin{array}{cc}
f'_{m_1} (M)  & 0 \\
0  & f'_{m_2}(M) \\
\end{array}
\right)  \rr,
\end{align}
where $f'_{m_1}$ and $f'_{m_2}$ are new polynomials of $M$ with degree $m_1$ and $m_2$, respectively.
This shows that the sources of BCOs, which were originally encoded in the coefficients of $g_{m_1m_2}(M)$, 
are actually redundant and can be absorbed into the redefinitions of the sources of 
the boundary preserving operators in $f_{m_1}$ and $f_{m_2}$. 
Thus, after the redefinitions, the partition function becomes independent of the sources of BCOs.
In terms of the original parametrization, this implies that there exist differentials 
$\nabla_n \; (n=1,2, \cdots, {\rm min}(m_1,m_2))$ such that they are given by 
linear combinations of the derivatives of the sources and satisfy 
$\nabla_n (F_{m_1m_2}) =0$. This is the simplest example of 
what we call the null identities.
These identities enable us to write the correlation numbers of BCOs in 
terms of those of boundary preserving operators.
We will present a general derivation of the null identities and show that
$\nabla_n$ can be constructed in such a way that the curvature is vanishing (i.e. $[\nabla_n, \nabla_l]=0$).
Then, we will discuss physical implications of the identities.

This paper is organized as follows. 
In Section 2, we derive the null identities.
In section 3 we  show some examples of the differentials $\nabla_n$ and the null identities,
and discuss physical implications of them. 
Section 4 is devoted to conclusion and
discussion on a possible extension to the cases where 
more than two boundary conditions are allowed.
We present the case with three boundary parameters in some detail.

\section{The null identities}

Under the double scaling limit of the one-matrix model, 
insertions of the matrix $M$ in the path integral 
can be replaced with insertions of a quadratic differential operator $Q$,
which acts on the space of eigenvalues of $M$ \cite{BDSS1989,D1989}.
Additive and multiplicative constants appear in this replacement: $M\rightarrow \epsilon Q+c$. 
For insertions of polynomials of $M$, these constants 
can be absorbed into renormalizations of the coefficients of 
the polynomials and the overall factors of the operators. 
After the renormalization, the perturbed partition function takes the form,
\be
\label{matrix:free:ene}
	F_{m_1m_2} = - \langle {\rm tr } \log R_{m_1m_2}(Q) \rangle,
	\quad
	R_{m_1m_2} (Q)=
	\left(\begin{array}{cccc}
	C_{m_1}(Q) & c(Q) \\  c(Q) & C_{m_2}(Q) 
	\end{array}\right),
\ee
where $C_{m_i}(Q)$ and $c(Q)$ are polynomials obtained by 
renormalizing $f_{m_i}$ and $g_{m_1m_2}$, respectively. They 
are written as
\bea
	C_{m_i}(Q)&=\prod_{k=1}^{m_i}\left(Q+a^{(i)}_k\right), \quad
	c(Q)=\sum_{n=0}^{d} c_{1+d-n} Q^{n},
\een
where $d=-1+{\rm min}\left(m_1,m_2\right)$ and $c_k$ real.
The coefficients of $C_{m_i} (Q)$ and $c(Q)$ correspond to 
the sources of boundary preserving and changing operators, respectively.
We will discuss this correspondence later in more detail
after we derive the null identities in the following.

By the formula ${\rm tr }\log R(Q) = \log {\rm det}R(Q)$, 
the perturbed partition function (\ref{matrix:free:ene}) can be written as the 
expectation value of the logarithm of ${\rm det}(R(Q))$.
As a polynomial of $Q$,  the degree of ${\rm det}(R(Q))$ is $m=m_1+m_2$ 
and it has $m$ independent coefficients.
However, the matrix $R(Q)$ has $m+d+1$ parameters in (\ref{matrix:free:ene}). 
Hence, $d+1$ parameters are redundant and those extra coefficients 
can be absorbed into redefinitions of the coefficients.
This implies that
there exist $d+1$ constraints on the partition function:
\be
\label{nullid2by2}
	\nabla_n F_{m_1m_2}=0,
\ee
which we refer to as null identities.
Here $n=1,2, \cdots , d+1$ and $\nabla_n$ are linear differential operators 
given by combinations of $\left\{ \frac{\p}{\p a^{(i)}_k}, 
\frac{\p}{\p c_n} \right\}$.
The differential operators $\nabla_n$ are specified by the condition
\be
\label{nabla def0}
	\nabla_n \left({\rm det} R_{m_1m_2}(x)\right)=0,
\ee
where $x$ is a formal parameter representing $Q$.
We express ${\rm tr}\, R_{m_1m_2}(x)$ as
\be
	\det  R_{m_1 m_2}(x)
	\equiv x^m+\sum_{k=1}^m \zeta_k\, x^{m-k}.
\label{def:zeta}
\ee
The operators $\nabla_n$ specified by \eqref{nabla def0} are 
equivalently defined by requiring the following conditions: For $ \forall  k\in\{1,2,\dots,m\}$
\be
	\nabla_n \, \zeta_k=0.
\label{nabla def}
\ee

A general solution to (\ref{nabla def}) can be constructed as follows. 
First, note that for each $c_n$
there should exists an independent differential operator
satisfying (\ref{nabla def}).
Then, we put an ansatz,
\begin{align}
	\nabla_n
	=\frac{\p}{\p c_n} + \tilde{\nabla}_n,
\label{ansatz for derivative}
\end{align}
where $\tilde{\nabla}_n$ is a linear differential operator consisting
of $\{a_k=a^{(1)}_k ;  a_{k+m_1}=a^{(2)}_k\}$. Specifically, 
$\tilde{\nabla}$ is written as 
\be
\tilde{\nabla}_n = \sum_{i,k} {\eta^{(n)}_k} \frac{\partial}{\partial a_k},
\ee
where $\eta^{(n)}_k$ are functions of $\{a_k; c_n\}$. 
The coefficients $\eta^{(n)}_k$ can be determined by requiring the conditions (\ref{nabla def}).
Let us introduce a set of variables,
\be
	\xi_i =  \sum_{1\leq j_1<j_2<\dots<j_i \leq m} a_{j_1} a_{j_2}\dots  a_{j_i}\, .
\label{def:xi}
\ee
Then, if $\zeta_k-\xi_k \ (\forall  k\in\{1,2,\dots,m\})$ has 
no explicit $a_i$-dependence (which is always the case for $2\times2$ block matrix
dealing with two boundary parameters), 
the differential operator is given by
\be
\label{null-for-2-by-2}
	\nabla_n
	=\frac{\p}{\p c_n}
	-\sum_{k=1}^m\frac{\p \zeta_k}{\p c_n}\frac{\p}{\p \xi_k}.
\ee

The conditions for the differential operators  \eqref{nabla def} allow
an ambiguity in the overall normalizations.
This ambiguity is fixed in \eqref{ansatz for derivative}
by setting the coefficients of the $c_n$-derivatives to be unity. 
This choice is very useful, since with this choice, the operators
mutually commute: $[\nabla_n, \nabla_l]=0$. 
This can be seen as follows. In general, $[\nabla_n, \nabla_l]$ is 
a linear differential operator. Since both of $\nabla_n$ and 
$\nabla_l$ satisfies \eqref{nabla def}, their commutator $[\nabla_n, \nabla_l]$ should also
satisfy \eqref{nabla def}. Then 
$[\nabla_n, \nabla_l]$ should be again given by a linear combination
of $\{\nabla_n \}$:
\begin{align}
[\nabla_n, \nabla_l]= \sum_k \alpha^{nlk}\nabla_k.
\label{lie algebra}
\end{align}
With the choice of \eqref{ansatz for derivative},
the left-hand side of 
(\ref{lie algebra}) does not contain $c_n$-derivatives,
while the right-hand side does. 
This means that $\alpha^{nlk}=0$ and thus $[\nabla_n, \nabla_l]=0$.

\section{Physical implications for correlation numbers}

The null identities \eqref{nullid2by2} provide important facts  
that any correlation numbers of boundary changing operators 
can be rewritten in terms of the correlation of boundary preserving operators. 
The possibility is due to the fusion rule between BCO operators. 

Let us consider the simplest case $ F_{11}=-\ll\log{\rm det}(R_{11}(Q))\rr $, 
where  $R_{11}(Q))$ is a  $2 \times 2$ matrix   
 \begin{align}
R_{11}(Q) = \matt[Q+a_1 ,  c  ,  c , Q+a_2],
\label{partition function for 1,1}
\end{align}
where $a_i$'s are cosmological constants of (1,1) boundaries
and assumed to take different values $a_1\neq a_2$.
The off-diagonal component $c$ 
couples to the boundary changing operator $B_{11}$
intertwining two different (1,1) boundaries
and produces  one null operator 
\begin{align}
	\nabla
		=\frac{\p}{\p c}+2c\frac{\p}{\p \xi_{2}},
\label{nabla for 1,1}
\end{align}
with $ \xi_1=a_1+a_2$ and $ \xi_2=a_1a_2$. 
We have the null identity 
$ \nabla^N F_{11}=0$   where $N$ is a positive integer. 
For $N=1$, the identity shows 
\be 
0= \left. \nabla F_{11}\right|_{c=0} = 
\left. \frac{\partial F_{11}}{\partial c} \right|_{c=0},
\ee
which is consistent with the fact that the one-point correlation of 
BCO is not allowed, since the boundary conditions contradict with each other.
For  $N=2$,  the two-point correlation of BCO
is given in terms of one-point boundary preserving correlation numbers:
\be 
\label{2BCO-11}
\langle B_{11} B_{11} \rangle
=  \left.\frac{\partial^2 F_{11} }{\partial c^2} \right|_{c=0} 
= -2 \langle I_2 \rangle  ,
\ee
where we define 
\be
 \langle {I_2}^k \rangle   =  \left.\frac{\partial^k F_{11} }{\partial \xi_2^k } \right|_{c=0}  .
 \ee
Using  $\p/\p\xi_2=- (1/a_{12})(\p/\p a_1-\p/\p a_2)$ 
with  $a_{12}= a_1 - a_2$, 
the result can be rewritten as 
  \be 
  \langle I_2 \rangle   
= \frac{ \langle  O_1  \rangle - \langle  O_2 \rangle }{a_{12} },
\ee 
where 
\be
\langle O_i  \rangle  =  \left.\frac{\partial  F_{11} }{\partial a_i } \right|_{c=0}
=\ll {\rm tr} \frac 1{Q+ a_i}  \rr.
\ee
The one-point correlation $\langle O_i  \rangle$
becomes $u^{1 / b^{2}} \cosh \left(\frac{\pi s_i}{b}\right)$ if 
one evaluates it at value $a_i=u \cosh \left(\pi b s_i \right)$,
where $u$ is a scale factor and $s_i$ a boundary parameter.

It is noted that the free energy  is given as 
\be
\label{F11-solution}
F_{11} = e^{-c^2 \frac{\p }{\partial \xi_2 }}  F_{11}^{(D)}(a_1,a_2),
\ee 
where $F_{11}^{(D)}$ is the $c$-independent part.
This shows that the 
cubic correlation of the BCO is absent and four-point correlation 
\be 
\langle B_{11} B_{11}  B_{11}  B_{11}\rangle 
= \left.\frac{\partial^4 F_{11} }{\partial c^4} \right|_{c=0} 
 =  12  \left.\frac{\partial^2 F_{11} }{\partial \xi_2^2} \right|_{c=0}
 =12 \langle I_2 I_2 \rangle .
\ee
In a similar manner, from null identities obtained by successive applications of $\nabla$,
one can find identities relating higher-point correlation numbers of BCOs with 
lower-point correlation numbers of boundary preserving operators.

One may look into a little complicated case:  BCO between   $(1,1)$ boundary
and $ (1,2)$ boundary.  
This can be investigated  using $F_{12}
	=-\ll\log {\rm det}(R_{12}(Q))\rr,$ 
where 
\begin{align}
	 R_{12}(Q) = \matt[Q+a_1 ,  c  ,  c ,  \left(Q+a_2\right)\left(Q+a_3\right)].
\end{align}
In this case also there is one off-diagonal parameter
which couples to BCO $B_{12}$.
The null operator is given as 
\begin{align}
	\nabla
	=\frac{\p}{\p c}+2c\frac{\p}{\p \xi_{3}},
\end{align}
with 
$\xi_3=\mu_1\mu_2\mu_3$
and provides a similar null identity as in between $ (1,1)$ boundaries:  $\nabla^N F_{12}=0$.
It is obvious that one has an alternative expression of the free energy as in  \eqref {F11-solution}
\be
\label{F12-solution}
F_{11} = e^{-c^2 \frac{\p }{\partial \xi_3 }}  F_{12}^{(D)}.
\ee 
In this case as well, correlation numbers with insertions of odd number of BCO $B_{12}$ are prohibited.
Two-point correlation is similarly given as in \eqref{2BCO-11}
\be 
\label{2BCO-12}
\langle B_{12} B_{12} \rangle 
= -2 \left.\frac{\partial F_{12} }{\partial \xi_3 } \right|_{c=0} = -2 \langle I_3 \rangle .
\ee
Here
$\langle I_3 \rangle$ is given in terms of one-point correlations of the boundary preserving operator $O_i$:
\be  
\langle B_{12} B_{12} \rangle  = 
-2\left(
\frac{\ll O_1\rr}{a_{21}a_{31}}+\frac{\ll O_2\rr}{a_{12}a_{32}}
+\frac{\ll O_3\rr}{a_{13}a_{23}}
\right),
\ee
with $a_{ij}:=a_i-a_j$.
It is noted that $(1,2)$ boundary condition is realized when $a_2=\mu_+$ and $a_3=\mu_-$
with $\mu_{\pm}=u \cosh \left(\pi b (s_2\pm ib)\right)$ and  $s_2$ real.
In this case one has $\left.\ll O_2\rr\right|_{a_2=\mu_+}=\left.\ll O_3\rr\right|_{a_3=\mu_-}
=-u^{1/b^2}\cosh(\pi b/s_2)$ and the two-point correlation of BCO becomes\footnote{
The evaluation at specific values of
boundary cosmological constants shall be indicated with subscript asterisk hereafter.}
\be
\langle B_{12} B_{12} \rangle_*  
=-
\frac{u^{\frac{1}{b^2}-2}\cosh \left(\frac{\pi s_{p}}{2 b}\right) \cosh \left(\frac{\pi s_{m}}{2 b}\right)}
{\sinh \left(\frac{\pi b\left(s_{p}+i b\right)}{2}\right) 
\sinh \left(\frac{\pi b\left(s_{p}-i b\right)}{2}\right) 
\sinh \left(\frac{\pi b\left(s_{m}+i b\right)}{2}\right) 
\sinh \left(\frac{\pi b\left(s_{m}-i b\right)}{2}\right)},
\ee
with $s_{p}=s_1+s_2$ and $s_{m}=s_1-s_2$ \cite{IR2010}.

Suppose we consider BCO between two different $(1,2)$ boundaries: 
$F_{22} 	=-\langle \log{\rm det} (R_{22}(Q))\rangle  $ where 
 \be 
	R_{22}(Q) = \matt[\left(Q+a_1\right) \left(Q+a_2\right),
			  c_1Q+c_2,  c_1Q+c_2, 
			  \left(Q+a_3\right)\left(Q+a_4\right) ].
\ee
The off-diagonal terms has two real parameters $c_1$ and $c_2$ 
and thus there are two independent  commuting null operators: 
\begin{align}
	\nabla_1&
	=\frac{\p}{\p c_1}+2\left(c_1\frac{\p}{\p \xi_{2}}+c_2\frac{\p}{\p \xi_{3}}\right),
	\\
	\nabla_2&
	=\frac{\p}{\p c_2}+2\left(c_1\frac{\p}{\p \xi_{3}}+c_2\frac{\p}{\p \xi_{4}}\right),
\end{align}
where $\xi_i$ is defined by \eqref{def:xi},
implying null identities $\nabla_1^{N_1} \nabla_2^{N_2}  F_{22}=0$.
The free energy can be written in the form
\be
F_{22} = e^{-c_1^2 \frac{\p}{\p \xi_{2}} - 2 c_1c_2 \frac{\p}{\p \xi_{3}} 
-c_2^2 \frac{\p}{\p \xi_{4}}}   ~F_{22} ^{(D)},
\ee
and therefore,
no correlations with odd number of BCOs $B_{22}^{(1)}$ and $B_{22}^{(2)}$,
which are associated with coupling constants $c_1$ and $c_2$, respectively, are allowed.
There are 3 kinds of two-point correlations 
\begin{align}
\label{2B22}
\langle B_{22}^{(1)}  B_{22}^{(1)}  \rangle &= -2 \langle I_2 \rangle
=-2\left(
\frac{a_1^2\ll O_1\rr}{a_{21}a_{31}a_{41}}+
\frac{a_2^2\ll O_2\rr}{a_{12}a_{32}a_{42}}+
\frac{a_3^2\ll O_3\rr}{a_{13}a_{23}a_{43}}+
\frac{a_4^2\ll O_4\rr}{a_{14}a_{24}a_{34}}
\right),
\\ 
\langle B_{22}^{(1)}  B_{22}^{(2)}  \rangle &= -2 \langle I_3 \rangle
=2\left(
\frac{a_1\ll O_1\rr}{a_{21}a_{31}a_{41}}+
\frac{a_2\ll O_2\rr}{a_{12}a_{32}a_{42}}+
\frac{a_3\ll O_3\rr}{a_{13}a_{23}a_{43}}+
\frac{a_4\ll O_4\rr}{a_{14}a_{24}a_{34}}
\right),
\\
\langle B_{22}^{(2)}  B_{22}^{(2)}  \rangle &= -2 \langle I_4 \rangle
=-2\left(
\frac{\ll O_1\rr}{a_{21}a_{31}a_{41}}+
\frac{\ll O_2\rr}{a_{12}a_{32}a_{42}}+
\frac{\ll O_3\rr}{a_{13}a_{23}a_{43}}+
\frac{\ll O_4\rr}{a_{14}a_{24}a_{34}}
\right),
\end{align}
where 
 $\langle B_{22}^{(i)}  B_{22}^{(j)}  \rangle
 =   {\partial^2 F_{22} }/{\partial c_i c_j } |_{c=0}  $
 and  $ \langle I_i \rangle =  {\partial F_{22}^{(D)} }/{\partial \xi_i  }  $.

To find BCO correlations between $(1,2)$ boundaries
we need to put correct parameterization of $a_i$'s:
 $a_{1,2}=u \cosh \left(\pi b (s_1\pm ib)\right)$,
$a_{3,4}=u \cosh \left(\pi b (s_2\pm ib)\right)$.
It is notable that $\langle B_{22}^{(1)} B_{22}^{(2)}\rangle_* \ne 0$.
One may find an orthogonal frame 
so that  $\langle \widetilde{B}_{22}^{(1)}  \widetilde{B}_{22}^{(2)}  \rangle_*  = 0 $,
where $\widetilde{B}_{22}^{(i)}$ is a new BCO associated with a new parameter $q_i$,
defined by $c_1Q+c_2 =   q_1 P_1 (Q)+q_2 P_0 $
where $P_i (Q)$ is an $i$-th order polynomial in $Q$:
$P_0=1$ and $P_1=Q-\frac{\langle B_{22}^{(1)}  B_{22}^{(2)}  \rangle_*}
{\langle B_{22}^{(2)}  B_{22}^{(2)}  \rangle_*}$
with
\be
\begin{aligned}
	\frac{\langle B_{22}^{(1)}  B_{22}^{(2)}  \rangle_*}{\langle B_{22}^{(2)}  B_{22}^{(2)}  \rangle_*}
	=-\frac{u\left(\cosh(\pi bs_1)+\cosh(\pi bs_2)\right)}{2\cos(\pi b^2)},
\end{aligned}
\ee
as given in \cite{BIR2010}.

One can extend the discussion to BCOs between $(1,m_1)$ and $(1,m_2)$ boundaries without any difficulties
using the null operator \eqref{null-for-2-by-2}.
It is noted that $(\p \zeta_k/\p c_n)$ has no $\xi_i$-dependence.
As a result, the free energy has no correlations with odd number insertions of BCOs.

\section{Conclusion and discussion}

In this paper, we considered correlation numbers of boundary 
changing and preserving operators in the
$(2,2p+1)$ minimal Liouville gravity on disk. 
In terms of the one-matrix model, we showed that those correlation
numbers satisfy some identities, called the null identities in 
this paper. These identities enable us to express correlation numbers
including boundary changing operators in terms of correlation numbers
with only boundary preserving operators.
This means that the correlation numbers of the boundary changing operators
can be determined from those of boundary preserving operators.
In addition, the null operator shows that the free energy has 
no cubic correlation of BCOs.

One may extend the matrix into $n \times n$ blocks
to describe correlation numbers with $n$ different boundary parameters:
\be
	F_{m_1m_2 \cdots m_n} = - \langle {\rm tr } \log R_{m_1 m_2 \cdots m_n}(Q) \rangle,
	\quad
	R_{m_1m_2 \cdots m_n} (Q)=
	\left(\begin{array}{cccc}
	C_1(Q) & c_{12}(Q) & \dots & c_{1n}(Q)  \\  c_{21}(Q) & C_2(Q) & \dots & c_{2n}(Q) \\ 
	\vdots & \vdots & \ddots &  \vdots \\  c_{n1}(Q) & c_{n2}(Q) & \dots & C_n (Q)
	\end{array}\right),
\ee
with $C_i(Q)$ and $c_{ij}(Q)$ respectively being
\bea
	C_{i}(Q)&=\prod_{k=1}^{m_i}\left(Q+a^{(i)}_k\right), \quad
	c_{ij}(Q)=c_{ji}(Q)=\sum_{n=0}^{d_{ij}} c^{(ij)}_{d_{ij}+1-n} Q^{n},
\een
where $d_{ij}=-1+{\rm min}\left(m_i,m_j\right)$.
The coefficients of $C_i (Q)$ and $c_{ij}(Q)$ are identified with
the sources of boundary preserving and changing operators.

Under this setup, as opposed to $2\times2$ block diagonal case,
there seems in general no explicit formula for mutually commuting 
differential operators $\nabla_n$ that satisfies
$\nabla_n\left(\det R_{m_1m_2 \cdots m_n} (x)\right)=0$.
However, still it is possible to find them under making ansatz \eqref{ansatz for derivative}
by requiring the conditions \eqref{nabla def}, where the parameters $\{\zeta_i\}$ and $\{\xi_i\}$
are understood as straightforward extensions of \eqref{def:zeta} and \eqref{def:xi}, respectively.
For example, with $3\times 3$ block matrix:
\begin{align}
	F_{111}
	=-\ll\log{\rm det}(R_{111}(Q))\rr,  \;\;\; 
	R_{111}(Q) = 
		\mat3[Q+a_1, c_3, c_2,
			 c_3, Q+a_2, c_1,
			 c_2, c_1, Q+a_3].
\end{align}
There are three commuting differential operators that provides 
the null identities:
\be
	\nabla_i
	=\frac{\p}{\p c_i}+2c_i\frac{\p}{\p \xi_2}+2e_i\frac{\p}{\p \xi_3}
	\qquad (i=1,2,3),
\ee
where 
$\xi_2=a_{1}a_{2}+a_{2}a_{3}+a_{3}a_{1}$,
$\xi_3=a_{1}a_{2}a_{3}$.
The coefficients $e_i$ $(i=1,2,3)$ are explicitly given by 
\be
\begin{aligned}
	e_1
	&=
	\frac{
	a_2a_{13}\,c_1{c_2}^2
	-a_3a_{12}\,c_1{c_3}^2
	+a_1a_{32}\,{c_1}^3
	+a_{32}a_{12}a_{13}\left(c_3c_2-a_1c_1\right)
	}
	{
	a_{32}{c_1}^2
	-a_{31}{c_2}^2
	+a_{21}{c_3}^2
	-a_{21}a_{31}a_{32}},
	\\
	e_2
	&=
	\frac{
	a_3a_{21}\,c_2{c_3}^2
	-a_1a_{23}\,c_2{c_1}^2
	+a_2a_{13}\,{c_2}^3
	+a_{13}a_{23}a_{21}\left(c_3c_1-a_2c_2\right)
	}
	{
	a_{32}{c_1}^2
	-a_{31}{c_2}^2
	+a_{21}{c_3}^2
	-a_{21}a_{31}a_{32}},
	\\
	e_3
	&=
	\frac{
	 a_1a_{32}\,c_3{c_1}^2
	-a_2a_{31}\,c_3{c_2}^2
	+a_3a_{21}\,{c_3}^3
	+a_{21}a_{31}a_{32}\left(c_2c_1-a_3c_3\right)
	}
	{
	a_{32}{c_1}^2
	-a_{31}{c_2}^2
	+a_{21}{c_3}^2
	-a_{21}a_{31}a_{32}},
\end{aligned}
\ee
which depends on $\xi_i$'s unlike in the $2\times 2$ case. 
As a result, the free energy,
satisfying the null identity $\nabla_1^{N_1}\nabla_2^{N_2}\nabla_3^{N_3}F_{111}=0$,
has non-vanishing cubic correlation of BCOs, for example,
\be
	\ll B^{a_2a_3}_{11}B^{a_3a_1}_{11}B^{a_1a_2}_{11}\rr
	=\left. \frac{\p^3 F_{111} }{\p c_1\p c_2\p c_3}\right|_{c=0}
	=-2\left({{a_{12}\ll O_3\rr+a_{23}\ll O_1\rr+a_{31}\ll O_2\rr}
 	\over{ a_{12}\,a_{23}\,a_{31} }}\right).
\ee

As one considers 
bigger size matrix with its block components of higher order polynomials,
their expression becomes more and more complicated,
still one can expect to find out the differential operators case-by-case.

\section*{Acknowledgments}

The work of G. I. was supported, in part, 
by Program to Disseminate Tenure Tracking System, 
MEXT, Japan and by KAKENHI (16K17679 and 19K03818).
The work of H. M. and C. R was partially supported by 
National Research Foundation of Korea grant number 2017R1A2A2A05001164.


\begin{thebibliography}{99}

\bibitem{Knizhnik1988}
V.~Knizhnik, A.~Polyakov, A.~B. Zamolodchikov, Fractal structure of 2d-quantum
  gravity, Mod. Phys. Lett. A3 (1988) 819.
\bibitem{BK1990}
E.~Brezin,  V.~Kazakov, Exactly Solvable Field Theories of Closed Strings,
Phys. Lett. B236 (1990) 144.
\bibitem{DS1990}
M.~Douglas, S.~Shenker, Strings in Less Than One-Dimension, 
Nucl. Phys. B335 (1990) 635.
\bibitem{GM1990}
D.~Gross, A.~Migdal, Nonperturbative Solution of the Ising Model on a Random Surface,
Phys.~Rev.~Lett. 64 (1990) 717. 
\bibitem{Ginsparg:1991bi}
For more references, see, for example, 
P.~H.~Ginsparg, Matrix models of 2-d gravity,
hep-th/9112013.
\bibitem{MSS1991}
G.~Moore, N.~Seiberg, M.~Staudacher, From loops to states in 2-d quantum
  gravity, Nucl. Phys. B362 (1991) 665--709.
\bibitem{BZ2009}
A.~A.~Belavin, A.~B.~Zamolodchikov,
{On correlation numbers in 2d
  minimal gravity and matrix models}, J.Phys.A 42 (2009) 304004.
\bibitem{BR2010}
A.~A.~Belavin, C.~Rim, Bulk one-point function on disk in one-matrix model,
  Phys. Lett. B687 (2010) 264--266.
 \bibitem{IR2010}
G.~Ishiki, C.~Rim, Boundary correlation numbers in one matrix model, Phys.  Lett. B694 (2010) 272.
\bibitem{FZZ2000}
V.~A.~Fateev, A.~B.~Zamolodchikov, Al.~B.~Zamolodchikov, 
{Boundary liouville field theory I. boundary state and boundary two-point function}
  (2000), arXiv:hep-th/0001012.
\bibitem{BDSS1989}
T.~Banks, M.~Douglas, N.~Seiberg, S.~Shenker, Microscopic and macroscopic loops
  in nonperturbative two-dimensional gravity, Phys. Lett. B238 (1990) 279.
\bibitem{D1989}
M.~Douglas, Strings in less than one-dimension and the generalized K-D-V
  hierarchies, Phys. Lett. B238 (1990) 176.
\bibitem{BIR2010}
J.~E.~Bourgine, G.~Ishiki, C.~Rim, Boundary operators in minimal liouville
  gravity and matrix models, JHEP 12 (2010) 046.


\end{thebibliography}
\end{document}